\documentclass[12pt,preprint]{aastex}

\slugcomment{Accepted by ApJ}

\begin{document}


\title{Long-term Monitoring on Mrk 501 for Its VHE $\gamma$ Emission and a Flare in October 2011}


\author{B.~Bartoli\altaffilmark{1,2},
 P.~Bernardini\altaffilmark{3,4},
 X.J.~Bi\altaffilmark{5},
 C.~Bleve\altaffilmark{3,4},
 I.~Bolognino\altaffilmark{6,7},
 P.~Branchini\altaffilmark{8},
 A.~Budano\altaffilmark{8},
 A.K.~Calabrese Melcarne\altaffilmark{9},
 P.~Camarri\altaffilmark{10,11},
 Z.~Cao\altaffilmark{5}
% ...
\\ (The ARGO-YBJ Collaboration)}


 \altaffiltext{1}{Dipartimento di Fisica dell'Universit\`a di Napoli
                  ``Federico II'', Complesso Universitario di Monte
                  Sant'Angelo, via Cinthia, 80126 Napoli, Italy.}
 \altaffiltext{2}{Istituto Nazionale di Fisica Nucleare, Sezione di
                  Napoli, Complesso Universitario di Monte
                  Sant'Angelo, via Cinthia, 80126 Napoli, Italy.}
 \altaffiltext{3}{Dipartimento di Matematica e Fisica ``E. De Giorgi" dell'Universit\`a del Salento, via per Arnesano, 73100 Lecce, Italy.}

 \altaffiltext{4}{Istituto Nazionale di Fisica Nucleare, Sezione di
                  Lecce, via per Arnesano, 73100 Lecce, Italy.}
 \altaffiltext{5}{Key Laboratory of Particle Astrophysics, Institute
                  of High Energy Physics, Chinese Academy of Sciences,
                  P.O. Box 918, 100049 Beijing, P.R. China. chensz@ihep.ac.cn}
 \altaffiltext{6}{Dipartimento di Fisica Nucleare e Teorica
                  dell'Universit\`a di Pavia, via Bassi 6,
                  27100 Pavia, Italy.}
 \altaffiltext{7}{Istituto Nazionale di Fisica Nucleare, Sezione di Pavia,
                  via Bassi 6, 27100 Pavia, Italy.}
 \altaffiltext{8}{Istituto Nazionale di Fisica Nucleare, Sezione di
                  Roma Tre, via della Vasca Navale 84, 00146 Roma, Italy.}
 \altaffiltext{9}{Istituto Nazionale di Fisica Nucleare - CNAF, Viale
                  Berti-Pichat 6/2, 40127 Bologna, Italy.}
 \altaffiltext{10}{Dipartimento di Fisica dell'Universit\`a di Roma ``Tor Vergata'',
                   via della Ricerca Scientifica 1, 00133 Roma, Italy.}
 \altaffiltext{11}{Istituto Nazionale di Fisica Nucleare, Sezione di
                   Roma Tor Vergata, via della Ricerca Scientifica 1,
                   00133 Roma, Italy.}
% ...

\begin{abstract}
As one of the brightest active blazars in both X-ray and very high energy $\gamma$-ray bands, Mrk 501 is very useful for physics associated with jets from AGNs. The ARGO-YBJ experiment is monitoring it for $\gamma$-rays above 0.3 TeV since November 2007. Starting from October 2011 the largest flare since 2005 is observed, which lasts to about April 2012. In this paper, a detailed analysis is reported. During the brightest $\gamma$-rays flaring episodes from October 17 to November 22, 2011, an excess of the event rate over 6 $\sigma$ is detected by ARGO-YBJ in the direction of Mrk 501, corresponding to an increase of the $\gamma$-ray flux above 1 TeV by a factor of 6.6$\pm$2.2 from its steady emission. In particular, the $\gamma$-ray flux above 8 TeV is detected with a significance better than 4 $\sigma$. Based on time-dependent synchrotron self-Compton (SSC) processes, the broad-band energy spectrum is interpreted as the emission from an electron energy distribution parameterized with a single power-law function with an exponential cutoff at its high energy end.
The average spectral energy distribution for the steady emission is well described by this simple one-zone SSC model. However, the detection of $\gamma$-rays above 8 TeV during the flare challenges this model due to the hardness of the spectra. Correlations between X-rays and $\gamma$-rays are also investigated.

\end{abstract}


\section{Introduction}
Blazars, including BL Lac objects and flat-spectrum radio quasars (FSRQs), are the most extreme subclass of active galactic nuclei (AGN). Most of the identified extragalactic $\gamma$-ray sources belong to this category. Their emission is believed to be dominated by non-thermal and strongly Doppler-boosted radiation from a relativistic jet of magnetized plasma which is aligned along our line of sight. The physical mechanism for the production of their $\gamma$-ray emission is still under debate. The leptonic models attribute the $\gamma$-ray emission to the inverse Compton scattering of the synchrotron (synchrotron self-Compton, SSC) or external photons (external Compton, EC) by the same population of relativistic electrons \citep{ghise98, dermer92,sikor94}, therefore an X-ray/$\gamma$-ray correlation is expected. The lack of strong emission lines in the radiation from BL Lac objects is taken as one of the evidences for a minor role of ambient photons (e.g., \citet[]{krawczy04}), and hence the SSC model is favored. The hadronic models attribute the $\gamma$-ray emission to proton-initiated cascades and/or proton-synchrotron emission in a magnetic field-dominated jet \citep{aharo00}. However, the tight X-ray and very high energy (VHE) $\gamma$-ray correlation and the very rapid $\gamma$-ray variability are taken as the strong challenges to models based on hadronic processes. Recently, a long-term continuous monitoring of Mrk 421 has been performed based on the ARGO-YBJ experiment and satellite borne X-ray detectors \citep{barto11}. According to this investigation, both the temporal and the spectral results generally favor the SSC model.
Even in the framework of the SSC model, the fundamental question referred to the origin of the flux and spectral variability, observed on timescales from minutes to tens of years, is still open.

Mrk 501 (z = 0.034)  was discovered with VHE emission by the Whipple collaboration \citep{quinn96}. It is one of the best-studied blazars with extensive studies on various timescales. In 1997, Mrk 501 went into a state with surprisingly high activity and strong variability and became more than a factor 10 brighter (above 1 TeV) than the Crab Nebula \citep{djann99,aharo99,ameno00}. The fastest $\gamma$-ray flux variability on a timescale of minutes was observed in 2005 \citep{albert07}. Significant spectral variability was detected with the harder spectrum  at the brighter states compared to the low-activity states \citep{albert07,ander09,accia11}. In the middle of 2009, a  multi-frequency observational campaign for 4.5 months on Mrk 501 was carried out with excellent energy coverage from radio to VHE $\gamma$-ray when  it underwent a low activity \citep{abdo11}.
Throughout the campaign, the source was sampled quite uniformly in all wavelength bands except for the VHE band, because the Cherenkov Telescopes cannot operate during non-optimal weather conditions or bright moonlight periods.
In October 2011, Mrk 501 underwent a strong flare detected by the MAXI satellite in X-rays \citep{sootome11} and by the ARGO-YBJ detector in VHE $\gamma$-rays \citep{barto11c}.

To understand the variability of emission and the underlying acceleration and radiation mechanisms in jets, continuous multi-wavelength observations from the X-ray to the VHE $\gamma$-ray band are crucial especially over a very long term. The broad band energy spectra could provide constraints on the parameters of the models.
The Cherenkov telescopes cannot constantly monitor AGNs because of their limited duty cycle and narrow  field of view (FOV).
The wide-FOV ARGO-YBJ detector, operated with high duty cycle ($> 85\%$), is more suitable for monitoring.
Working at energies above 300 GeV, ARGO-YBJ extends the multi-wavelength survey carried out by the satellite borne X-ray detector $Swift$ and the GeV $\gamma$-ray detector $Fermi$-LAT.
Particularly, the spectral energy distribution (SED) of Mrk 501 is covered without any gap from 100 MeV to 10 TeV. All the measurements would set strong constrains on the model of the emission of AGNs. \par
In this paper, we report on the multi-wavelength view of the emission from Mrk 501 from August 2008 to April 2012, including the  average spectra during  quasi-steady  and   flaring periods.

\section{The ARGO-YBJ experiment and data analysis}
The ARGO-YBJ experiment, located at the Cosmic Ray Laboratory of Yangbajing
(Tibet, P.R. China) at an altitude of 4300 m a.s.l., is the result of a collaboration among Chinese and Italian institutions and is designed for VHE $\gamma$-ray astronomy and cosmic ray observations. The detector consists of a single layer of Resistive Plate Chambers (RPCs), which
are equipped with charge readout
strips (6.75 cm $\times$ 61.80 cm  each). The logical OR of signals from eight neighboring strips
constitutes the pixel (called a  ``pad'') for triggering and timing
purposes. Each RPC is read with ten pixels.
130 clusters (each composed of 12 RPCs) are installed to form a carpet
of about 5600 m$^{2}$ with an active area of $\sim$93\%. This central
carpet is surrounded by 23 additional clusters (``guard ring''). The total area of the
array is  110 m $\times$ 100 m.
The ARGO-YBJ detector is operated by requiring the number of fired pads ($N_{pad}$) to be
at least 20 within 420 ns on the entire carpet.
The trigger rate is 3.5 kHz with a dead time of 4\%. The angular resolution, pointing accuracy and stability of
the ARGO-YBJ detector array have been thoroughly tested by
measuring the shadow of the Moon in cosmic rays \citep{barto11b}. The absolute energy scale uncertainty is less than 13\% for all measured cosmic ray showers \citep{barto11b}.
More details about the detector and
the RPC performance can be found in \citep{aielli06,aielli09a,aielli09b,aielli09c}.

For the multi-wavelength investigation together with $Fermi$-LAT, the data collected by ARGO-YBJ after August 2008 when $Fermi$ was launched are used.
The total effective observation time is  1179.6 days. To achieve a good angular resolution,
events with zenith angle
less than 50$^{\circ}$ are used, and further selection criteria \citep{barto11,barto12} are applied. The total number of events after filtering is $1.86\times$10$^{11}$ for this
work. No $\gamma$/hadron discrimination is applied. The opening angle $\psi_{70}$, which contains 71.5\% of the events from a point-like source, is
1.36$^{\circ}$ for events with $N_{pad}>60$.
In order to remove the effect of the cosmic-ray anisotropy, the method described in \citep{barto11,barto12} was applied.
The significance of the excess is estimated using the \cite{li83} method. With this data analysis, the significance of the excess observed
from the direction of the Crab Nebula is 17 standard deviations ($\sigma$) in 3.5 years, which
indicates that the 3.5-year cumulative 5 $\sigma$ sensitivity of ARGO-YBJ
has reached 0.3 Crab unit for point sources \citep{cao11}.

\section{Results}
\subsection{Light curves}
\begin{figure}
\epsscale{0.8}
\plotone{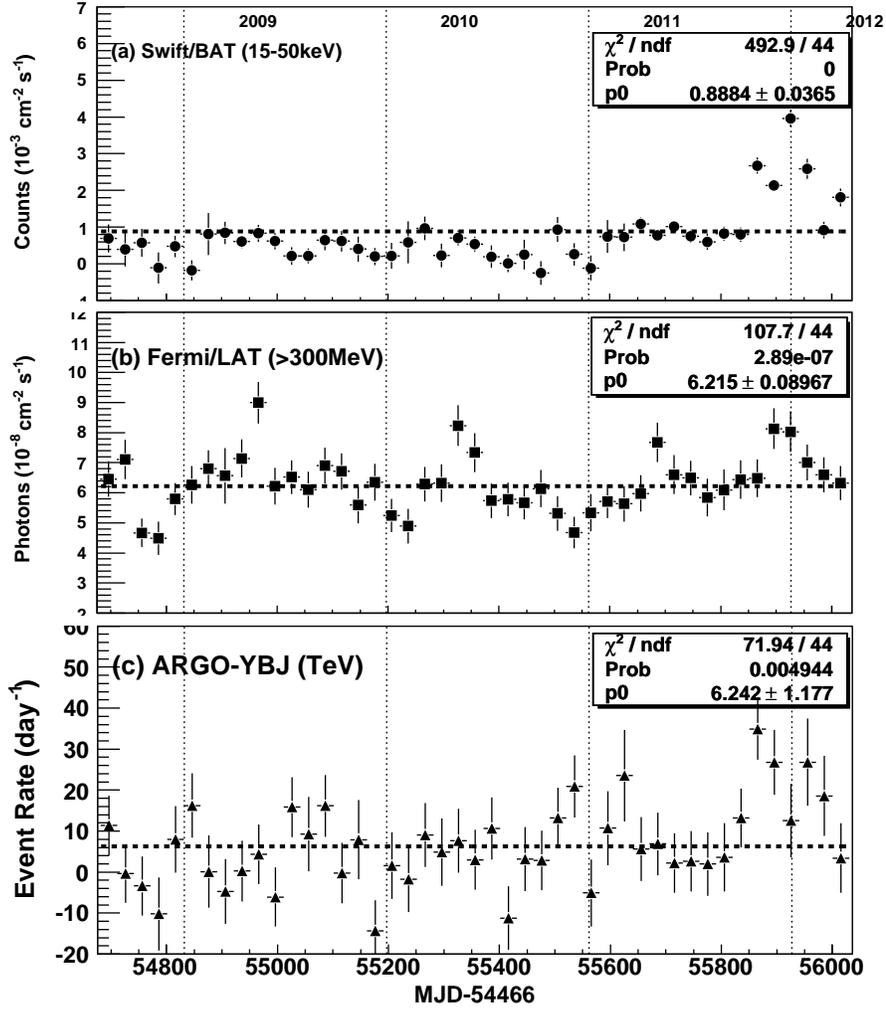}
\caption{Light curves of Mrk 501 with 30-day bins. The vertical bars represent the 1-$\sigma$
uncertainties. The horizontal dashed lines
and the legends (for all curves) show the results of a fit with a constant value to the data set. }
\label{fig1}
\end{figure}

\begin{figure}
\epsscale{0.8}
\plotone{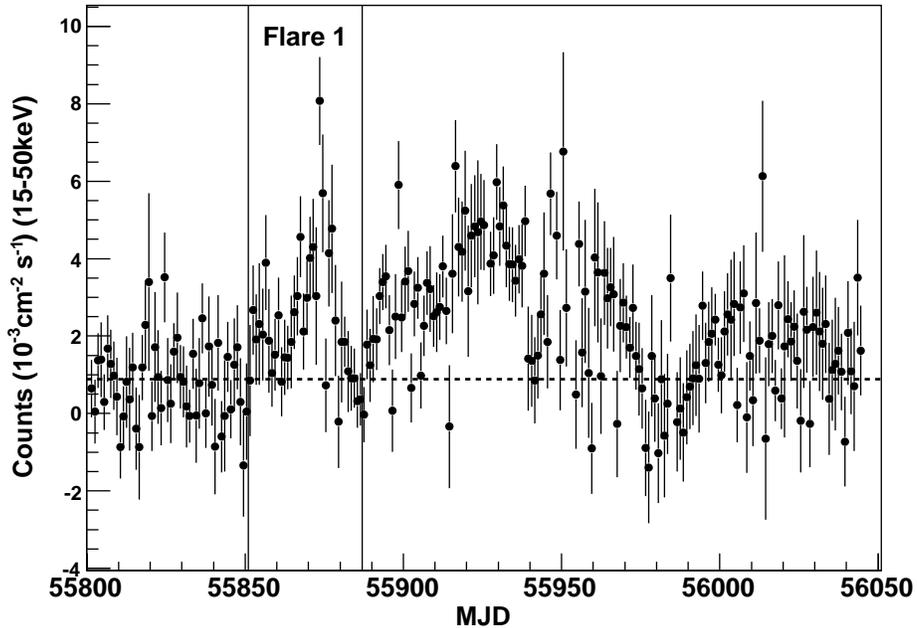}
\caption{Light curves of Mrk 501 with 1-day bins obtained with the Swift/BAT instrument at 15-50 keV.  The vertical bars represent the 1-$\sigma$ uncertainties. The vertical lines show the start and the end of flare 1. The horizontal dashed line is the same as those in Fig.1.}
\label{fig2}
\end{figure}

The daily flux from Mrk 501 at energy 15$-$50keV provided by $Swift$/BAT  \footnote{ Transient monitor results provided by the $Swift$/BAT team:
http://heasarc.gsfc.nasa.gov/docs/swift/ results/transients/weak/Mrk501/.} is publicly available and used in this work.
The light curve from August 2008 to April 2012 is shown in panel (a) of Figure~\ref{fig1} with a bin size of 30 days. The best fit with a constant value for the light curve is (8.9$\pm$0.4)$\times10^{-4}$ counts cm$^{-2}$ s$^{-1}$ with a $\chi^2$ of 492.9 for 44 degrees of freedom (ndf).
A significant feature is the flare at the end of 2011 with the flux being enhanced by a factor about 4.
Without data during the flaring period, the best fit with a constant value for the light curve is (6.0$\pm$0.4)$\times10^{-4}$ counts cm$^{-2}$ s$^{-1}$ with a $\chi^2/ndf$ of 69.3/38.
This result could be an evidence for a small X-ray variability before this flaring period.
Zooming in the $Swift$/BAT light curve at 15$-$50 keV shown in Figure~\ref{fig2}, the large flare started in October 17, 2011 (MJD=55851) and decreased to a low-activity state around the average level on November 22 (MJD=55887)(flare 1 hereafter). Afterwards, Mrk 501 became increasingly active for a longer period up to about April 2012. Its brightest flaring episode is in November 8, 2011, during the flare 1 period.

The $Fermi$-LAT data were analyzed using the ScienceTools\footnote{http://fermi.gsfc.nasa.gov/ssc/}.
The light curve is generated using aperture photometry.
The panel (b) of Figure~\ref{fig1} shows the flux at energies greater than 0.3 GeV that is contained within a 2-degree cone centered on Mrk 501, which is the 68\% containment angle of the reconstructed incoming photon direction for normal incidence.
A fit with a constant value yields a $\chi^2/ndf$ of 107.7/44 indicating a moderately variable behavior consistent with the X-ray analysis.
The Discrete Correlation Function (DCF), computed as
prescribed by \cite{edelson88}, for the BAT/LAT data points shown in
Figure~\ref{fig1} is DCF = 0.63$\pm$0.26 for a time lag of zero, which is greater than the previously measured DCF = 0.32$\pm$0.22 \citep{abdo11}.
Since the significance in both analysis is less than 2.5$\sigma$, only minor correlations are observed.
During the X-ray flaring period, the GeV $\gamma$-ray flux increased above the long-term average, but this flux increase is not significant.
Therefore, the light curve does not indicate a significant correlation with the X-ray data during the flare.

The light curve in the TeV $\gamma$-ray range detected by ARGO-YBJ is shown in panel (c) of Figure~\ref{fig1}. During the X-ray flare, the flux of TeV $\gamma$-rays also increases. A fit with a constant emitting rate yields a $\chi^2/ndf$ of 71.9/44,
while the $\chi^2/ndf$ is reduced to 42.59/38 by simply excluding the data during the X-ray flares. The TeV $\gamma$-ray flare 1 was clearly detected as a counterpart of that in the X-ray band, and the count rate increase is by a factor about 6.
The DCF for the BAT/ARGO-YBJ data points shown in
Figure~\ref{fig1} is  0.85$\pm$0.36 for a time lag of zero, while the DCF for the LAT/ARGO-YBJ is 0.44$\pm$0.31.

\subsection{Photon Energy Spectra}
\begin{figure}
\plottwo{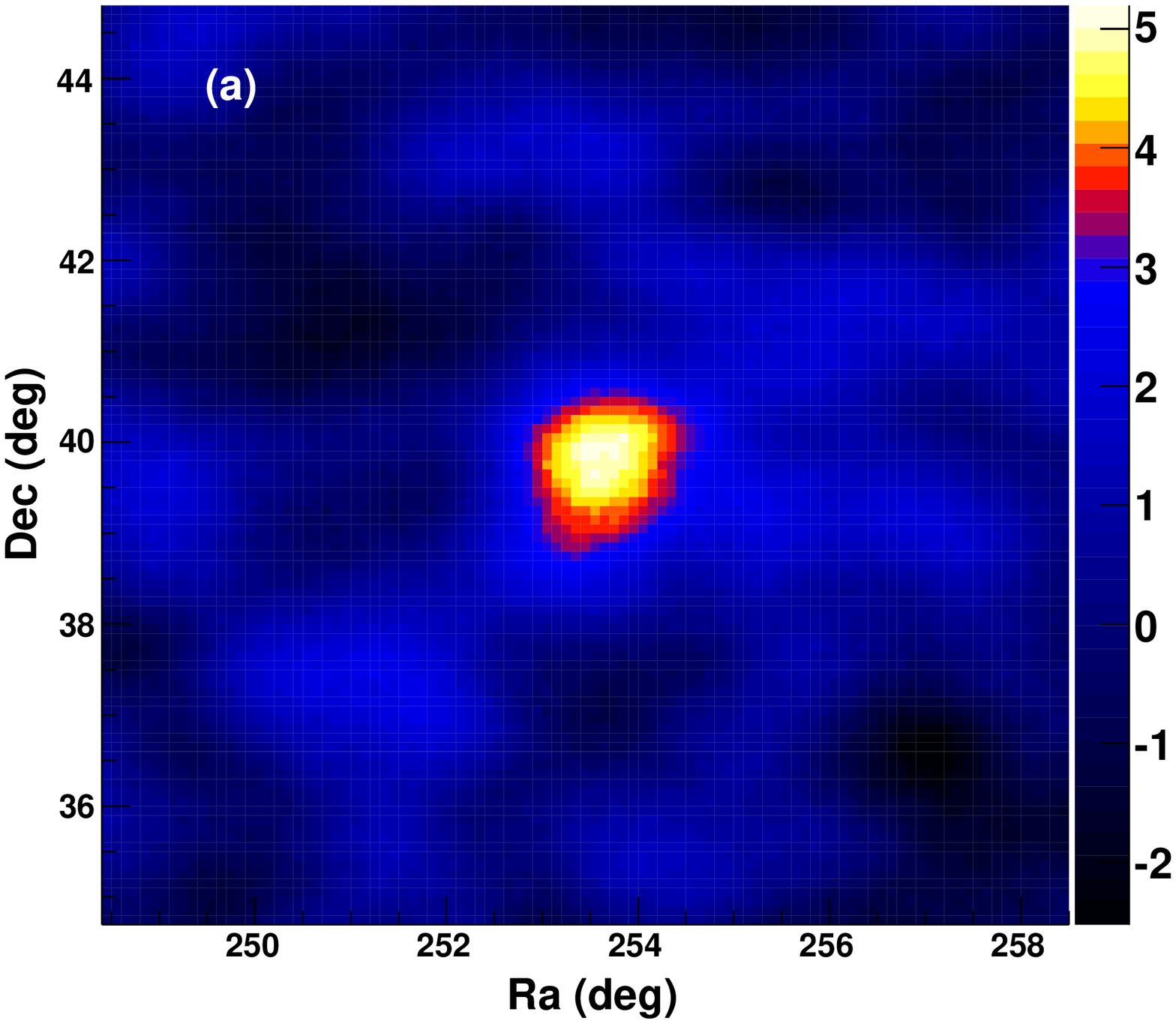}{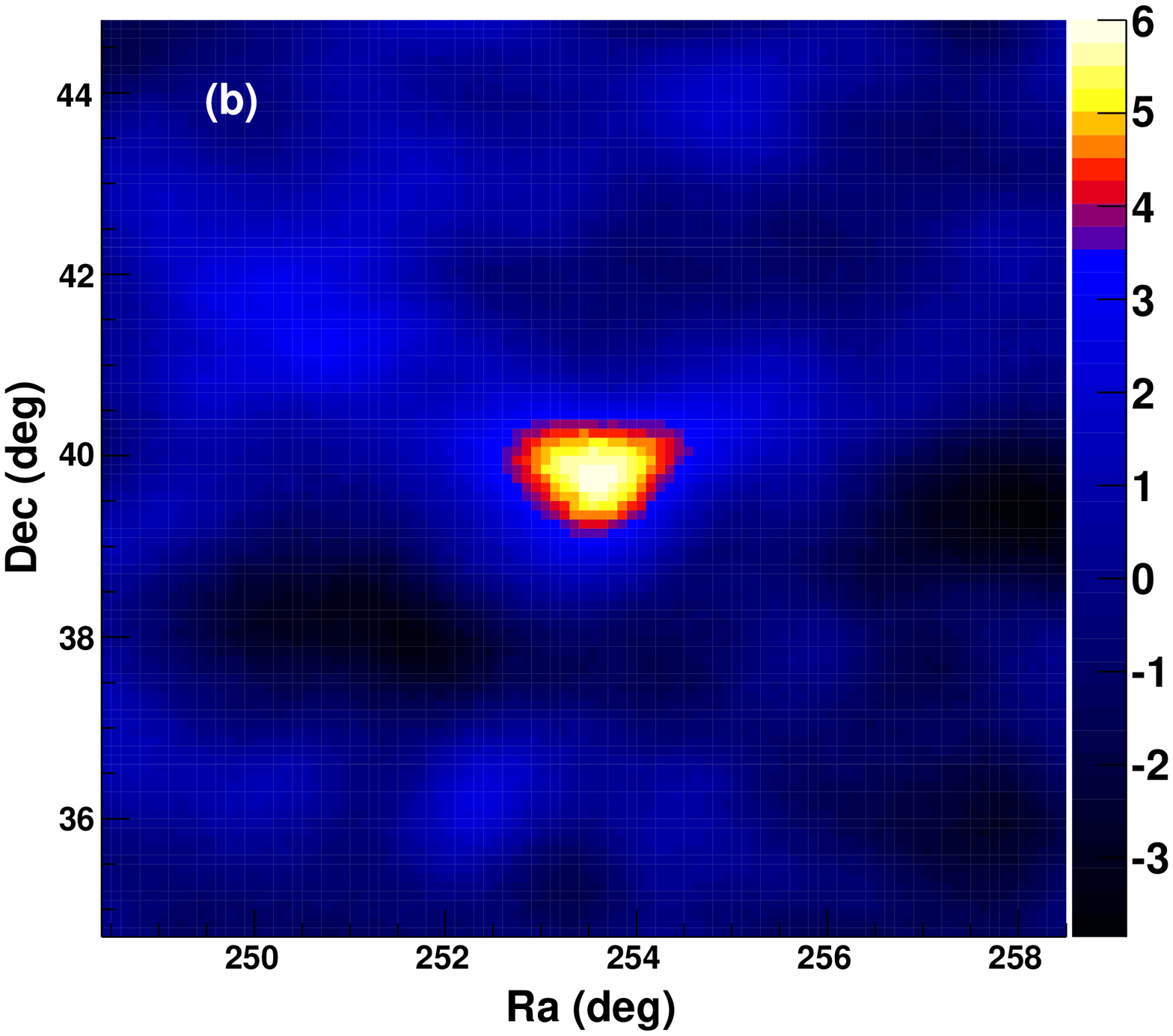}
\caption{Significance map of the Mrk 501 region: (a) shows the statistical significance in standard deviations in the period from MJD 54683 to 55850; (b) shows the statistical significance of the flaring period from MJD 55851 to 55887. }
\label{fig3}
\end{figure}

To investigate the evolution of the spectra, their time-averages during the long-term quasi-steady state from August 5, 2008 to October 16, 2011 and during flare 1 are estimated separately. The integrated flux from Mrk 501 observed by ARGO-YBJ has a statistical significance of 5 $\sigma$ (see panel (a) of Figure~\ref{fig3}). During flare 1, the flux from Mrk 501 was detected by ARGO-YBJ with 6.1 $\sigma$ significance (see panel (b) of Figure~\ref{fig3}), corresponding to an increase of the $\gamma$-ray flux above 1 TeV by a factor of 6.6 from its steady emission.
Events with $N_{pad}>60$ are used for both panels of Figure~\ref{fig3}.
For comparison, all the spectra are shown in Figure~\ref{fig4}, where an average spectrum over 4.5 months was adapted from \citep{abdo11}. Without significant activity observed in the last 3 years before flare 1, the average spectra over 4.5 months could approximately represent the average spectra over 3 years. All the other spectra presented in Figure~\ref{fig4} are estimated as described in the following.

\subsubsection{Swift: X-ray}
The $Swift$/XRT is a focusing X-ray telescope with energy range from 0.2 to 10 keV. In Windowed Timing mode during flare 1, XRT data in four time windows are available from HEASARC \footnote{http://heasarc.gsfc.nasa.gov/}, and the exposure time is about 1 ks in each window.
The XRT data set was processed with the XRTDAS software package (v.2.6.0) following the standard recommendations. The XRT average spectrum in the 0.5$-$10 keV energy band
was fitted using the XSPEC package (v.12.7.0). We adopted a power-law
model for the photon-flux spectral density, with an absorption hydrogen-equivalent column
density fixed to the Galactic value in the direction of
the source, namely $1.56 \times 10^{20} $ $cm^{-2}$ \citep{kalber05}. We obtain the spectrum of $(0.0407 \pm 0.0007) \times \left(\frac{E}{1keV}\right)^{-1.824 \pm 0.022}$
${\rm ph}\, {\rm cm}^{-2}\,{\rm s}^{-1}\,{\rm keV}^{-1}$.

\subsubsection{ $Fermi$-LAT: HE $\gamma$-rays}
The LAT data from a region centered on Mrk 501 with radius of $10^{\circ}$ was used to estimate the spectrum. The data analysis was performed following the standard recommendations by using an unbinned maximum-likelihood method to estimate the source SED.
We adopted a power-law model for the energy range from 0.1 to 300 GeV. During the long-term period, Mrk 501 was detected with a Test Statistic (TS)
value of 5118.2 ($\sim 71.5 \sigma$).  We obtain the spectrum $(2.300 \pm 0.068) \times 10^{-12}\left(\frac{E}{1844.3MeV}\right)^{-1.774 \pm 0.021}$ ${\rm ph}\, {\rm cm}^{-2}\,{\rm s}^{-1}\,{\rm MeV}^{-1}$.  This is consistent with the result reported by the LAT collaboration using the data collected during the first 24 months \citep{abdo12}. The spectrum was found to be very stable in the GeV band in spite of a moderate variability over a long period.
During flare 1, the TS was 329.1 ($\sim 18.1 \sigma$) and the spectrum was $(3.53 \pm 0.46) \times 10^{-12}\left(\frac{E}{1844.3MeV}\right)^{-1.640 \pm 0.084}$ ${\rm ph}\, {\rm cm}^{-2}\,{\rm s}^{-1}\,{\rm MeV}^{-1}$. Compared with the long-term result, the flux increased slightly.

\subsubsection{ ARGO-YBJ: VHE $\gamma$-rays}
The VHE $\gamma$-ray spectrum was estimated using a distribution of the excess in the number of events as a function of $N_{pad}$.
 A widely used procedure, that is described in \citep{barto11}, was followed. In this procedure, the spectrum of Mrk 501 was assumed to be a power law.
The ARGO-YBJ detector response has been taken into account using G4argo \citep{guo10}. The simulated events are sampled in the energy range from 10 GeV to 100 TeV.
We define six intervals with $N_{pad}$
of 20$-$59, 60$-$99, 100$-$199, 200$-$499, 500$-$999 and $\geq$1000. The best fit gives a
differential flux(${\rm cm}^{-2}\,{\rm s}^{-1}\,{\rm TeV}^{-1}$)
\begin{equation}
(1.92 \pm 0.44) \times 10^{-12}\left(\frac{E}{2TeV}\right)^{-2.59 \pm 0.27}
\end{equation}
for the long-term period, corresponding to 0.312$\pm$0.076 Crab units above 1 TeV. The median energies of the six intervals are  0.45, 0.89, 1.4, 2.8, 5.6 and 11 TeV, respectively. Both the flux and the spectral index are similar to those obtained by VERITAS and MAGIC during the low-activity state, as presented in \citep{abdo11}. In particular, the ARGO-YBJ data show a smooth extension of the $Fermi$-LAT results, as evident in Figure~\ref{fig4}.
During flare 1, the differential flux (${\rm cm}^{-2}\,{\rm s}^{-1}\,{\rm TeV}^{-1}$)  is
\begin{equation}
(2.92 \pm 0.52) \times 10^{-12}\left(\frac{E}{4TeV}\right)^{-2.07 \pm 0.21}
\end{equation}
corresponding to 2.05$\pm$0.48 Crab units above 1 TeV, that is a factor 6.6$\pm$2.2 compared with its long-term steady state.
The median energies of the six intervals are 0.89, 1.1, 1.8, 3.5, 7.1 and 14 TeV, respectively.
Only the statistical error is quoted here, and the systematic uncertainty in the flux measurement is estimated to be $\lesssim$30\% \cite{aielli10}.

\section{Discussion}

\subsection{Spectra Corrected for the EBL Absorption}
\begin{figure}
\epsscale{0.8}
\plotone{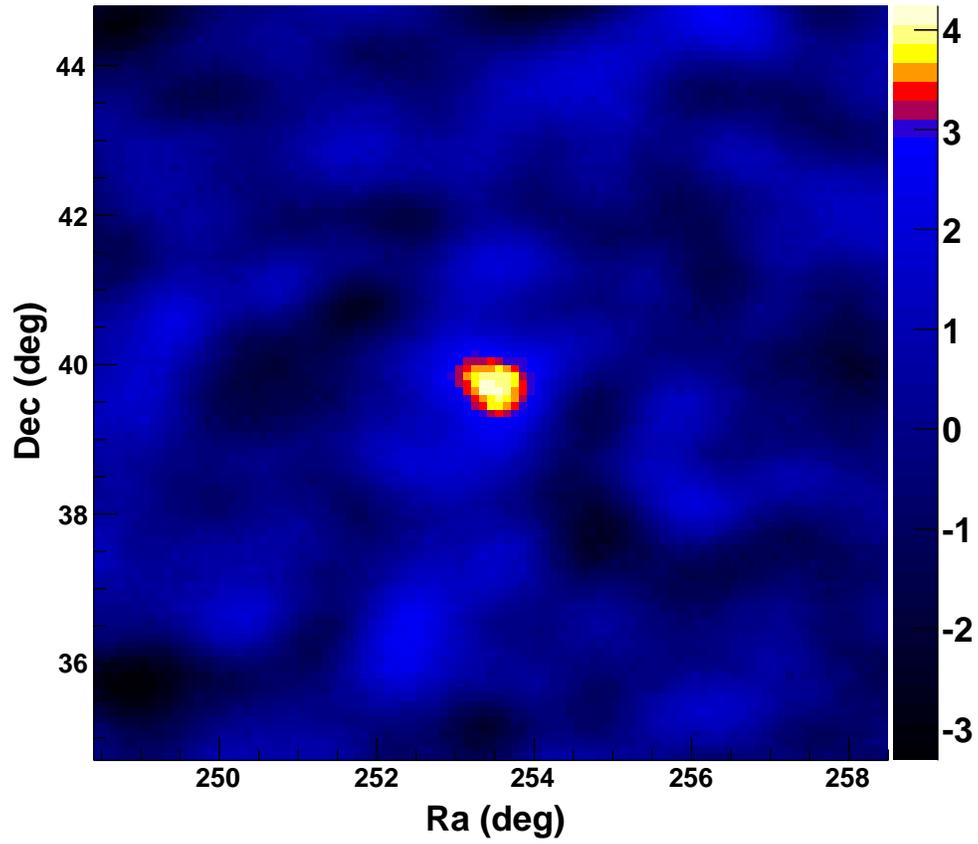}
\caption{Significance map of the Mrk 501 region during the flaring period from MJD 55851 to 55887. Events with $N_{pad}>500$ are used, whose corresponding median energy is 8.4 TeV. }
\label{fig5}
\end{figure}

During flare 1, $\gamma$-rays with a median energy of 8.4 TeV are also observed with a significance greater than 4$\sigma$, as shown in Figure~\ref{fig5}. $\gamma$-rays with such a high energy from Mrk 501 have not been detected since the 1997 flare. The SED at energies above 0.9 TeV is harder than those observed during the flares in 1997 \citep{aharo01} and in June 30, 2005 flare \citep{albert07}, as shown in Figure~\ref{fig4}, although the spectral indices are consistent if the statistical error is taken into account. Mrk 501 is a nearby source, so we do not expect a
significant absorption of its intrinsic source spectrum due to Extragalactic
Background Light (EBL) at energies below 1 TeV, while the absorption at higher energies is still considerable.
Therefore, it is useful to test different EBL models assuming a minimum intrinsic photon spectral index.
A natural minimum spectral index is 1.64, constrained by the spectrum in the GeV band,
since that at higher energies should be steeper.

\begin{figure}
\epsscale{0.8}
\plotone{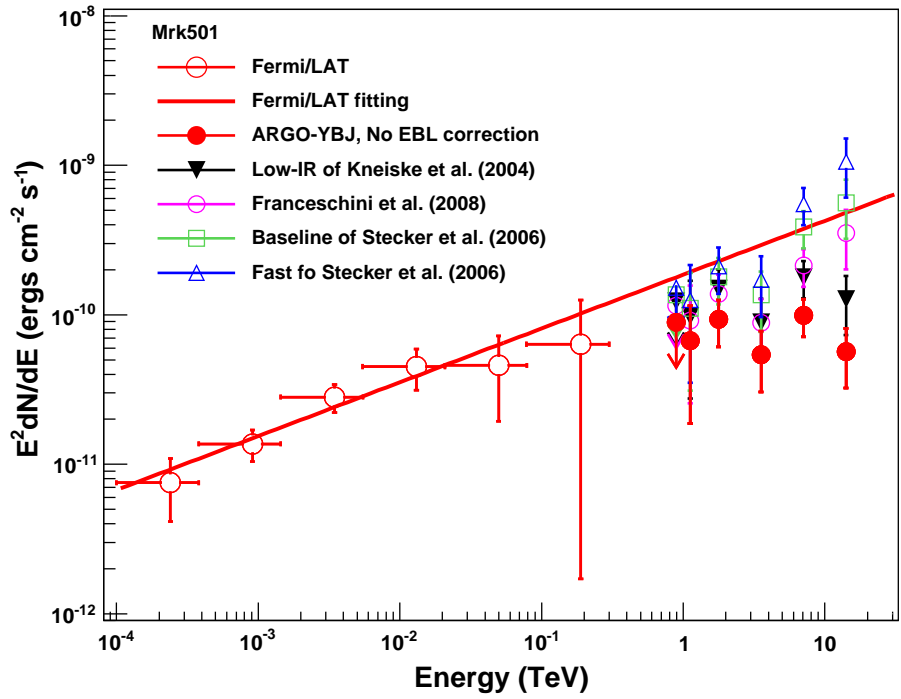}
\caption{Spectra of Mrk 501 during the flaring period from MJD 55851 to 55887. The spectrum from 0.1 GeV to 300 GeV is observed by $Fermi$-LAT. The solid line is the fit to the $Fermi$-LAT data using a power law. The spectrum observed by ARGO-YBJ has been corrected for
EBL absorption using four different EBL models. Details can be found in the text. }
\label{fig6}
\end{figure}

\begin{figure}
\epsscale{0.8}
\plotone{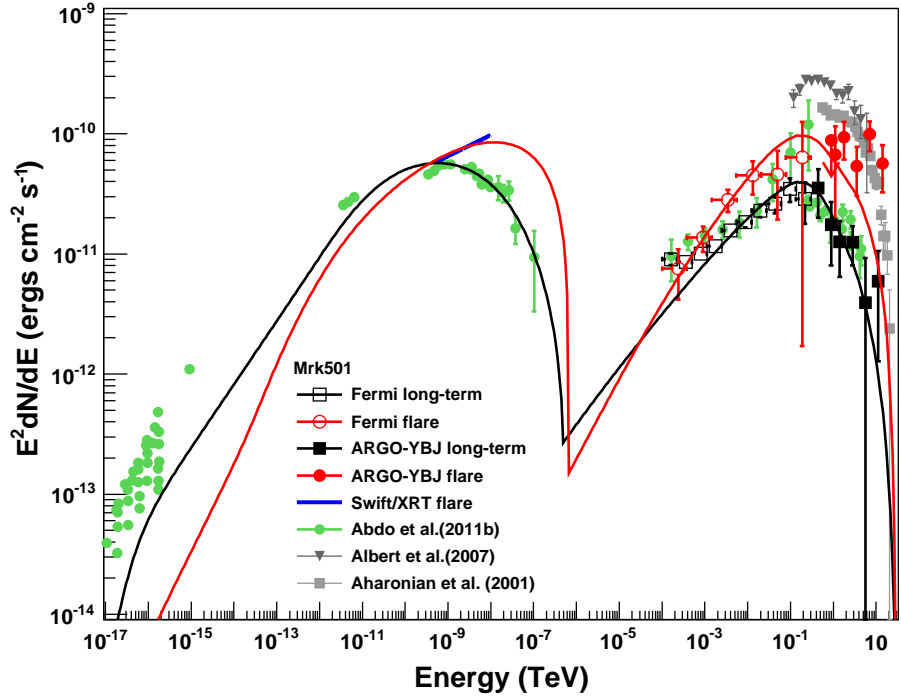}
\caption{Spectral energy distribution of Mrk 501.
The solid lines
show the best fits to the data with a one-zone SSC model, with the
best-fit parameters listed in Table 1.
The black curve corresponds to a SED model describing the long-term averaged
data, while the red curve describes the flaring data.}
\label{fig4}
\end{figure}

Here we use four kinds of models with different flux levels of the EBL, among many models. Assuming a single power law for the VHE flux, the indices of the derived unabsorbed spectra using different models are:(1)$1.80_{-0.29}^{+0.26}$ for the ``low-IR" model proposed by \cite{kneis04}, which gives a similar result compared with that obtained using the EBL model of \cite{aharo06}, according to \cite{albert07};
(2) $1.45_{-0.42}^{+0.36}$ for the model of \cite{franc08}, which is widely used to correct the VHE SED of extragalactic sources; (3,4)$1.30_{-0.38}^{+0.34}$ and $1.11_{-0.41}^{+0.37}$ for the baseline and fast evolution models  proposed by \cite{stecker06,stecker07}, respectively.
For comparison, all the unabsorbed spectra are shown in Figure~\ref{fig6}.
The spectral indices obtained using models (2), (3) and (4) exceed the minimum spectral index boundary of 1.64, however they are consistent with this limit if the statistical error is taken into account. It has to be noted that models (3) and (4) have been excluded with higher significance by previous tests carried out around 1 TeV \citep{aharo06,georg10} and tens of GeV \citep{abdo09,abdo10}. Since our data extend to about 10 TeV, the corresponding EBL photon energy is substantially lower. The EBL model with the minimum absorption is used when modeling the SED in the following section.

\subsection{Modeling of the overall SED}
The long-term averaged SED, especially the continuous measurement of the second component in the energy range 0.1 GeV-10 TeV obtained in this work, provides us a robust baseline for an insight into the underlying physics in Mrk 501.
Over the very wide radio-VHE energy range, we fitted a
one-zone SSC model proposed by \citet[][]{MaK95} \citep[see
also][]{MaK97,yanget08} to the SED. There are a few free parameters to be determined in this model, including the
Doppler factor $\delta=1/[\Gamma(1-\beta\cos\theta)]$, the spherical
blob radius R, the magnetic field strength $B$, the electron spectral
index $s$, the electron maximum Lorentz factor $\gamma_{\rm max}$, and the
electron injection compactness $l_e=\frac{1}{3}m_ec\sigma_{\rm
T}R^2\int_1^\infty d\gamma(\gamma-1)Q_e$. In order to determine the Doppler factor and the injection compactness, further parameters must be determined, i.e.,
the Lorentz factor $\Gamma$, the speed of the blob $c\beta$ and
the Lorentz factor of electrons $\gamma$. Moreover,
$\sigma_{\rm T}$ is the Thomson cross section,
 $\theta$ the angle between the direction of motion of the blob
and the observer's line of sight, and $Q_e$ the electron spectrum at injection, which is assumed to be a power law cutting off at $\gamma_{\rm max}$, with a normalization factor $q_e$,
$Q_e=q_e\gamma^{-s}\exp(-\gamma/\gamma_{\rm max})$.
This model  is different from other one-zone SSC models  which introduce
more free parameters using stationary injected electron spectra with a double power-law function \citep{tavec01,ander09} or even triple power-law function \citep{abdo11} with an exponential cutoff at their high energy end.
A full time-dependent evolution of the electron and photon spectra based on this model was simulated for a given injection spectrum of electrons.
The low-IR model proposed by \cite{kneis04} is used to take into account the
absorption of $\gamma$-rays in the EBL when modeling the SED.

The best fit to the long-term SED is shown
in Figure~\ref{fig4}, with the corresponding parameters given in Table 1.
General agreement between the model and the data is achieved with this simple one-zone SSC model.
The parameters in the model are found in general agreement with those found in the previous analysis on the similar source Mrk 421 \citep{barto11}.
The best fit for the flare 1 is also shown in Figure~\ref{fig4}, with the corresponding parameters given in Table 1.
The highest-energy data points of the SED ($>6$ TeV) during flare 1 cannot be well reproduced by only modifying the parameters.
It is important to point out that the X-ray spectrum becomes harder during the flare. The peak energy is shifted to about 10 keV during flare 1 from about 1 keV in the quasi-steady state.
The shift of the X/$\gamma$-ray peak to higher energies during flares is a common feature which has been reported many times (e.g. \citep{albert07,ander09,accia11}). However, the detection of $>$6 TeV $\gamma$-rays with an energy flux similar to that at 1 TeV is unusual. In the framework of the SSC model, it is difficult to reproduce such high energy $\gamma$-rays as detected by ARGO-YBJ. Since the $\gamma$-rays above 1 TeV are typically produced in the Klein-Nishina regime, their rate should be strongly suppressed. The radiation mechanism during flares may be different from that in the quasi-steady state.
Different radiation mechanisms, such as more complex SSC models or hadronic processes (e.g. in \citet{abra12}), are needed to improve the understanding of the flaring phenomena.

\section{SUMMARY}
We have presented a continuous long-term monitoring of Mrk 501 from August 2008 to April 2012.
Both the flux and the spectral index are consistent with those obtained
by VERITAS and MAGIC during a 4.5 month long multifrequency campaign
\citep{abdo11}.
Combining the observations by the ARGO-YBJ experiment with the space-borne experiments $Swift$ and $Fermi$,
the investigation was performed over a wide energy range from 0.5 keV up to 10 TeV (a value higher than the maximum in \citep{abdo11}).
Using all the data covering various energy bands during the quasi-steady phase of the blazar, its SED is fitted with a simple one-zone SSC model assuming a single power-law with exponential cutoff for the electron spectrum at injection.
The model parameters are found in agreement with those resulting from previous analyses for various AGNs, indicating that similar radiation mechanisms are in action.
A strong flare of the blazar in the VHE region was observed by ARGO-YBJ in October 2011, while no Cherenkov Telescope could observe Mrk 501 during this period.
It is well determined to be a counterpart of the X-ray flare in the same time
period, giving for the BAT/ARGO-YBJ data points a DCF of 0.85$\pm$0.36
for a time lag of zero.
On the contrary, there is no significant increase of the flux at energies around 1 GeV. Remarkably, $\gamma$-rays with energies above 8 TeV are detected, which did not happen since the 1997 flare.
The spectral shape obtained with the data in the GeV/TeV energy region during
the flare favors the ``low-IR" EBL model of \cite{kneis04}, while a simple one-zone SSC
model with a single power-law  electron spectrum at injection is not able to reproduce the spectral shape at the highest energies
(above 6 TeV).

\acknowledgments
 This work is supported in China by NSFC (No.10120130794),
the Chinese Ministry of Science and Technology, the
Chinese Academy of Sciences, the Key Laboratory of Particle
Astrophysics, CAS, and in Italy by the Istituto Nazionale di Fisica
Nucleare (INFN).

We are grateful to Yupeng Chen for his help in estimating the
X-ray spectrum using the $Swift$/XRT data.
We also acknowledge the essential supports of W.Y. Chen, G. Yang,
X.F. Yuan, C.Y. Zhao, R. Assiro, B. Biondo, S. Bricola, F. Budano,
A. Corvaglia, B. D'Aquino, R. Esposito, A. Innocente, A. Mangano,
E. Pastori, C. Pinto, E. Reali, F. Taurino and A. Zerbini, in the
installation, debugging and maintenance of the detector.


\clearpage

\begin{deluxetable}{cccccccc}
\tablecolumns{7} \tablewidth{0pc} \tablecaption{Best-Fit Parameters
in the SSC Model }
 \tablehead{
\colhead{Flux Level} & \colhead{$\gamma_{max}$}   &
\colhead{$l_{e}$} & \colhead{$B$ (G)} & \colhead{$R$ (cm)}    &
\colhead{$\delta$} & \colhead{$s$}}

\startdata
Long-term &$3\times10^6$ &  $1.8\times10^{-5}$ &   0.07& $ 3\times10^{16}$&  12&  1.95 \\
Flare 1 &$8\times10^6$ &  $8\times10^{-5}$ &   0.10& $ 1\times10^{16}$&  10&  1.6 \\
\enddata
\end{deluxetable}

\clearpage

\clearpage

\end{document}